\documentclass[
pre,
twocolumn,
showpacs,
 amsmath,
 amssymb,
floatfix,
]{revtex4-1}

\usepackage{graphicx}
\usepackage{dcolumn}% Align table columns on decimal point
\usepackage{color}
\usepackage{amsmath}
\usepackage{bm}
\usepackage{epsfig}

\newcommand{\Tref}{T_{\rm{L}}}
\newcommand{\ci}{{\bm{c}_i}}

\begin{document}

\title{{Particles-on-Demand for Kinetic Theory}}

\author{B. Dorschner }
\altaffiliation[Present address: ]{California Institute of Technology, Pasadena, CA 91125, USA}
\author{F. B\"{o}sch}
\author{I. V. Karlin}\thanks{Corresponding author}
\email{karlin@lav.mavt.ethz.ch}
\affiliation{
  Department of Mechanical and Process Engineering, ETH Zurich, 8092 Zurich, Switzerland 
}
	
\date{\today}
	
\begin{abstract}
A novel formulation of fluid dynamics as a kinetic theory with tailored,  on-demand constructed particles 
removes any restrictions on Mach number and
temperature as compared to its predecessors, the lattice
Boltzmann methods and their modifications.  
In the new kinetic theory, discrete particles are determined by a rigorous limit process which avoids ad hoc assumptions about their velocities.
Classical benchmarks for incompressible and compressible flows demonstrate that the proposed discrete-particles kinetic theory
opens up an unprecedented wide domain of applications for computational fluid dynamics.
\end{abstract}
	
%\pacs{}
%\keywords{Suggested keywords} %Use showkeys class option if keyword display desired

\maketitle

\noindent Kinetic theory of Boltzmann and Maxwell, as the fundamental link between
particles' picture of flowing matter and a continuum projection thereof, has
been a valuable source of ideas in fluid dynamics. This especially concerns the
lattice Boltzmann method (LBM)
\cite{frisch1986lattice,mcnamara1988use,higuera1989lattice,qian1992lattice},
a modern approach to the simulation of complex flows. 
LBM is a recast of fluid mechanics into a kinetic theory for the populations
of designer particles $f_i(\bm{x}, t)$, with simple rules of propagation on a
space-filling lattice formed by discrete speeds $\mathcal{C}=\{\bm{c}_i$, $i=1,\dots, Q\}$, in
discrete-time~$t$, and relaxation to a local equilibrium $f_i^{\rm eq}(\bm{x}, t)$  
at the nodes $\bm{x}$. 
LBM witnessed burgeoning growth in applications and becomes the method of
choice for complex fluid dynamics problems such as turbulence \cite{Ansumali2017},
wetting-dewetting transition \cite{Sbragaglia2006}, 
microfluidics \cite{Kunert2007,Hyvaeluoma2008}, microemulsions \cite{Benzi2009} and
hemodynamics \cite{Thiebaud2014}, to mention a few; recent reviews can be found in
\cite{Aidun2010,Krueger2017,Succi2017book}.\\
However, a critical look at LBM reveals major limitations: all practical LBM models are severely restricted in 
flow speed and temperature range.
While these restrictions may be traded for deeply subsonic, slow flows, even
then insufficient isotropy and lack of Galilean invariance impede simulations \cite{Qian1993}. Moreover,
the said limitations become eventually insurmountable for compressible flows~\cite{alexander1993lattice,guo2007thermal,he1998novel,mcnamara1995stabilization,shan1998discretization}.
It may be argued that LBM
has reached its natural limits with the simulation of quasi-incompressible flows, and a different discrete kinetic theory is needed for important fields such as combustion and aerodynamics.\\
In this Letter, we demonstrate that eventually all physical limitations of the LBM are
removed once the discrete kinetic theory is formulated using tailored rather than fixed particles' velocities at every space location and every time instance. The new fully explicit realization outperforms LBM by many orders in terms of flow speed and temperature. This opens door to kinetics-based simulations of fluid dynamics which were not possible before.\\
We begin with a clarification: LBM interprets the discrete speeds $\ci$ as particles' velocities, $\bm{v}_i^{\rm L}=\bm{c}_i$.
On the contrary, here we understand $\ci$ as {\it peculiar velocities} \cite{ChapmanCowling}, relative to a reference frame velocity $\bm u$ and a temperature $T$. Henceforth,  particles' velocities are defined as
\begin{align}
\label{eq:peculiar}
\bm{v}_i=\sqrt{\theta}\bm{c}_i+\bm{u},
\end{align}
where $\theta =T/ \Tref$ is the temperature reduced by the lattice temperature $\Tref$, a constant which is 
known for any set of discrete speeds $\mathcal{C}$ \cite{chikatamarla2006entropy}.
According to (\ref{eq:peculiar}), LBM amounts to setting a global reference frame \textquotedblleft at
rest," $\bm{u}=\bm{0}$, and choosing the fixed temperature $T=T_{\rm{L}}$ for all particles. 
Here, we rather follow the interpretation (\ref{eq:peculiar}) where the reference frame velocity and temperature are kept so far undetermined, and we are going to find optimal values for $\bm{u}$ and $T$, as presented in detail below.\\
By specifying the frame velocity and temperature in~\eqref{eq:peculiar},
one sets the reference frame (or gauge) $\lambda=\{\bm{u},T\}$ for the discrete
velocities.
LBM corresponds to the standard gauge $\lambda_{\rm L}=\{\bm{0},\Tref\}$.
We denote   
%$f^{\bm{u}}=\left( f^{\bm{u}}_1, \dots, f^{\bm{u}}_Q\right)$ 
$f^{\lambda}=\left( f^{\lambda}_1, \dots, f^{\lambda}_Q\right)^\dagger$ as
the vector of populations relative to the gauge
$\lambda$.
The transform of the populations to another gauge $\lambda'= \{\bm{u}',T'\}$ is facilitated by
matching $Q$ linearly independent moments ($m,n$ are integers;
%running from $0$ to $q-1$; 
$D=2$ to ease notation),
\begin{align}\label{eq:moments}
M^{\lambda}_{mn}=\sum_{i=1}^{Q}f_i^{\lambda}(\sqrt{\theta}{c}_{ix}+{u}_x)^m 
                                            (\sqrt{\theta}{c}_{iy}+{u}_y)^n.
\end{align}
{Let us use a short-hand notation for a linear map of populations into} {moments $M^{\lambda}$ (\ref{eq:moments}), $M^{\lambda}=\mathcal{M}_{\lambda}f^{\lambda}$, where $\mathcal{M}_{\lambda}$ is the $Q\times Q$ matrix of the linear map.}
The matching condition for the moments in both gauges $\lambda$ and $\lambda'$ reads,
\begin{align}\label{eq:match}
M^{\lambda}=M^{\lambda'}.
\end{align}
In other words, the moments of the populations are independent of the choice of a gauge.
Moments matching condition (\ref{eq:match}) implies that populations are transformed from one gauge to another
with the transfer matrix $\mathcal{G}_{\lambda}^{\lambda'}$,
\begin{align}
\label{eq:transform}
f^{\lambda'}=\mathcal{G}_{\lambda}^{\lambda'}f^{\lambda}=\mathcal{M}^{-1}_{\lambda'}\mathcal{M}_{\lambda}f^{\lambda}.
\end{align}
Finally, we introduce a reconstruction formula for populations at 
any point $\bm{x}$ at time $t$:
\begin{align}
\label{eq:interpolation}
\tilde{f}^{\lambda}(\bm{x},t)=\sum_{s=1}^{k}a_s(\bm{x}-\bm{x}_s)\mathcal{G}_{\lambda_s}^{\lambda}f^{\lambda_s}(\bm{x}_s,t),
\end{align}
where $\lambda_s=\{\bm{u}(\bm{x}_s,t), T(\bm{x}_s,t)\}$ are the gauges at the
collocation points $\bm{x}_s$, at time $t$, and $a_s$ are
interpolation functions (standard Lagrange polynomials below; $k$ determines the order). 
Note that the reconstruction formula (\ref{eq:interpolation}) enforces
populations at collocation points to be treated in a specified gauge $\lambda$
through the transform (\ref{eq:transform}).

\noindent
We now present the discrete kinetic theory in an optimal local gauge.
Introducing the time step $\delta t$, evaluation of the populations at the
monitoring point $\bm{x}$ at time $t$ involves the propagation and the
collision steps, mediated by the gauge transform.

\noindent
{\bf Propagation}. Semi-Lagrangian advection is  performed first,
using the reconstruction formula (\ref{eq:interpolation}) at the departure point of
characteristic lines, $\bm{x}-\bm{v}^0_i\delta t$, 
\begin{align}
\label{eq:advection}
f_i^{\lambda_0}=
\tilde{f}_i^{\lambda_0}\left(\bm{x}-\bm{v}_i^0\delta t,t-\delta t\right),
\end{align}
where the characteristic directions $\bm{v}_i^0$  (or discrete velocities, cf.\
Eq.\ (\ref{eq:peculiar})) are set relative to a seed gauge
$\lambda_0=\{\bm{u}_0,T_0\}$. 
For the latter, {it is convenient to} choose flow velocity and temperature at the monitoring point
$\bm x$ at time $t-\delta t$:
\begin{align}\label{eq:seed_gauge}
%\bm{v}_i=\bm{c}_i+\bm{u}(\bm{x},t-\delta t),
\bm{u}_0&=\bm{u}(\bm{x},t-\delta t),\\
T_0&=T(\bm{x},t-\delta t), \label{eq:seed_gauge_T}
\end{align} 
yielding
\begin{equation}\label{eq:vi}
\bm{v}^0_i=\sqrt{\theta_{0}}\bm{c}_i+\bm{u}_0,
\end{equation} 
with $\theta_0=T_0/ \Tref$.
Since $\bm{u}_0$ and $T_0$ are known from the previous time step, the populations
(\ref{eq:advection}) are determined unambiguously in this {\it predictor}
propagation step. %\\

\noindent
With the populations (\ref{eq:advection}), the density, momentum and
temperature are evaluated at the monitoring point
using discrete velocities (\ref{eq:vi}):
\begin{align}
\label{eq:conservation_rho}
\rho_1&=\sum_{i=1}^Qf_i^{\lambda_0},
%\left(\bm{x},t-\frac{\delta t}{2}\right)
\\
\label{eq:conservation_rhou}
\rho_1\bm{u}_1&=\sum_{i=1}^Q
%\left(\bm{x},t-\frac{\delta t}{2}\right)
\bm{v}_i^0
f_i^{\lambda_0},\\
\label{eq:conservation_T}
D\rho_1 T_1+\rho_1\|\bm{u}_1\|^2&=\sum_{i=1}^Q \|\bm{v}^0_i\|^2f_i^{\lambda_0}.
\end{align}
This defines the {\it corrector} gauge $\lambda_1=\{\bm{u}_1, T_1\}$ at the
monitoring point, and advection (\ref{eq:advection}) is executed anew with the
updated velocities, $\bm{v}^1_i=\sqrt{\theta_{1}}\bm{c}_i+\bm{u}_1$, to get corrected
post-propagation populations $f^{\lambda_1}_i$. The predictor-corrector
process is iterated until convergence, with the limit values,
\begin{align}
&\rho(\bm{x},t)=\lim\limits_{n\to\infty}\rho_n,\label{eq:density}\\
&\bm{u}(\bm{x},t)=\lim\limits_{n\to\infty}\bm{u}_n, \label{eq:flow}\\
&T(\bm{x},t)=\lim\limits_{n\to\infty}T_n,\label{eq:T}\\
&f_i^{\lambda(\bm{x},t)}=\lim\limits_{n\to\infty}{f}_i^{\lambda_n},\label{eq:precollision}
\end{align} 
defining the density (\ref{eq:density}), the flow velocity (\ref{eq:flow}), the
temperature (\ref{eq:T}) and the pre-collision  populations
(\ref{eq:precollision}) at the monitoring point $\bm{x}$ at time $t$.
{Note that, by construction, the limit gauge $\lambda(\bm{x},t)=\{\bm{u}(\bm{x},t),T(\bm{x},t)\}$ is the co-moving reference frame
in which the discrete particle's velocity (\ref{eq:peculiar}) is defined by the {values} of the flow velocity and of the temperature at the monitoring point.}

\noindent {\bf Collision}: 
In the co-moving reference frame, the local equilibrium populations are defined by
the density only,  
\begin{align}\label{eq:equilibrium}
f^{\rm eq}_i=\rho W_i, 
\end{align}
where the weights $W_i$ are known for any discrete speeds set $\mathcal{C}$ \cite{chikatamarla2006entropy}, see also Appendix~\ref{app:equilibrium}. 
[Note that, in the standard LBM context, populations (\ref{eq:equilibrium}) would be identified as local equilibrium \textquotedblleft at zero flow velocity $\bm{u}=\bm{0}$".]
Hence, pre-collision populations (\ref{eq:precollision}) are transformed to
post-collision as 
\begin{align}
\label{eq:postcollision}
f_i\left(\bm{x},t\right)=f^{\lambda(\bm{x},t)}_i
%+\omega\left[\rho(\bm{x},t)W(\bm{x},t)-f^{\bm{u}(\bm{x},t)}\right]
+2\beta\left[\rho(\bm{x},t)W_i-f^{\lambda(\bm{x},t)}_i\right],
\end{align}
for the Bhatnagar-Gross-Krook (BGK) collision model. 
{The} relaxation parameter $\beta$ is related to the kinematic viscosity {by} $\nu=T(1/2\beta-1/2)\delta t$.
By fixing the temperature and canceling the energy corrections
(\ref{eq:conservation_T}), one arrives at the isothermal version of the proposed kinetic theory.
Comments are in order here:

\noindent
{(i) In LBM, particles (represented by discrete velocities) are fixed once and for all with the identification
$\bm{v}_i^{\rm L}=\bm{c}_i$. Then the local equilibrium acquires
non-invariant dependence on the flow velocity and temperature which leads to errors once $\bm{u}\ne\bm{0}$ and $T\ne T_{\rm L}$. 
{Accumulation of these errors is also the primary source of numerical instabilities when the plain BGK collision model is used in LBM.}
On the contrary, the new representation of kinetics creates \textquotedblleft optimal particles" (or optimal discrete velocities), specific to each monitoring point at a given time (see propagation step) so that the equilibrium (\ref{eq:equilibrium}) \textquotedblleft seen"
by the populations becomes {\it exact}. Hence, this new representation is, in
principle, restricted neither in the flow speed nor in the
range of temperature variation. Error-free equilibrium can also result in unconditional numerical stability when using the BGK model. Below, we shall probe all this with benchmark simulations.
}\\
(ii) 
If the standard gauge $\lambda_{\rm L}$ is adopted,
then the transfer matrix $\mathcal{G}$  is dropped in 
(\ref{eq:interpolation}), and advection (\ref{eq:advection}) becomes 
$f_i=\tilde{f}_i\left(\bm{x}-\bm{c}_i\delta t,t-\delta t\right)$. 
The latter, together with finite element reconstruction, was used in a recent
semi-Lagrangian LBM (SLLBM) \cite{Kraemer2017}.  SLLBM is not
restricted to space-filling lattices and was realized on body-fitted
unstructured meshes \cite{DiIlio2018}, an obvious advantage if turbulent flow simulations are
concerned.
Present algorithm fully retains this crucial feature.\\
The standard two-dimensional  nine-speeds set $D2Q9$ was used in all simulations below and the BGK collision (\ref{eq:postcollision}) was implemented for both isothermal and compressible flow. The transfer matrix was found in closed form and is presented in 
Appendix~\ref{app:transfer},
% \cite{supplement}, 
together with the reconstruction formula realization.
The LBM time step $\delta t=1$ was used in all simulations.
\begin{figure}[t]
	\includegraphics[width=0.45\textwidth]{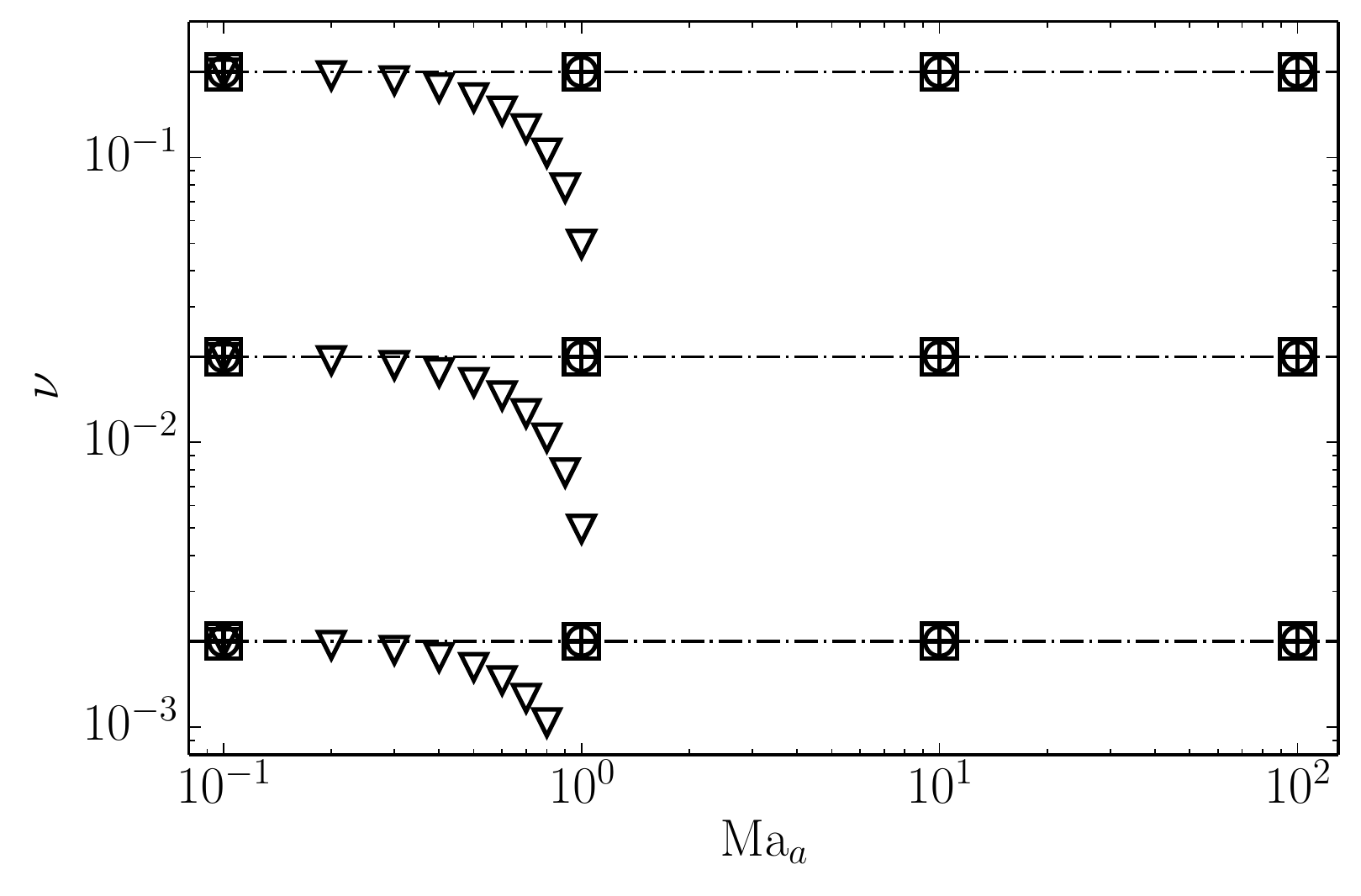}
    \caption{{Kinematic viscosity from decaying $\pi/4$-tilted shear wave with $D2Q9$, at
      various advection Mach numbers ${\rm Ma}_{\rm a} = u/\sqrt{T}$. Lattice temperature $T_{\rm L}=1/3$.
      Lines: imposed theoretical values $\nu=0.2$, $\nu=0.02$, $\nu=0.002$; Symbol: present method at fixed temperature $T_{\rm L}/2$ (cross), $T_{\rm L}$ (circle) and 
      $2T_{\rm L}$ (square);
      Triangle: LBGK \cite{qian1992lattice} at  $T_{\rm L}$.}}
	\label{fig:visc}
\end{figure}

\noindent First, we measured kinematic viscosity at isothermal conditions.
The 
decay of plane shear
wave with initial profile $u_\xi(\xi,\eta) = A \sin(2 \pi \xi/L)$, $A=0.05$,
in transverse direction and advection $u_\eta(\xi,\eta) = {\rm Ma}_{\rm a} \sqrt{T}$ in the wave-vector (longitudinal) direction was studied.
The wave vector was 
rotated {by $\pi/4$} with respect to the standard Cartesian $x$-axis and periodic boundary conditions were applied in both longitudinal and transverse directions.
This tilted-wave setup is standard to probe isotropy and Galilean invariance \cite{Qian1998,Hazi2006}: kinematic viscosity should not depend on the advection Mach number ${\rm Ma}_{\rm a}$.
An equidistant mesh with resolution $L=200\sqrt{2}$ in longitudinal direction was used.
Kinematic viscosity was measured by least square fit of exponentially decaying function.
In Fig.\ \ref{fig:visc}, the kinematic viscosity is shown for various temperatures, in a wide range of advection speeds. 
It is apparent that the results of the present formulation are in excellent agreement with theoretical prediction, for advection Mach numbers even as high as ${\rm Ma}_{\rm a} = 100$, and are independent of temperature.
This is in sharp contrast to the standard lattice BGK (LBGK) \cite{qian1992lattice} which
shows lack of Galilean invariance already at ${\rm Ma}_a \gtrsim 0.1$.
While the latter failure of LBM has been long known \cite{Qian1998,Hazi2006},  it is striking that a mere reformulation of the {\it same} kinetic model in the optimal gauge extends validity by at least three orders of magnitude in terms of flow speed.
Note that, since the temperature can be set at a high value, and not only at $T=T_{\rm L}$ as in the LBM, the quasi-incompressible flow simulations can be performed at realistic Mach numbers with the present method. This was used in the Green-Taylor vortex simulation at ${\rm Ma}\sim 10^{-3}$ which confirmed second-order convergence, see Appendix~\ref{app:convergence}. 
\begin{figure}[t]
	\includegraphics[width=0.45\textwidth]{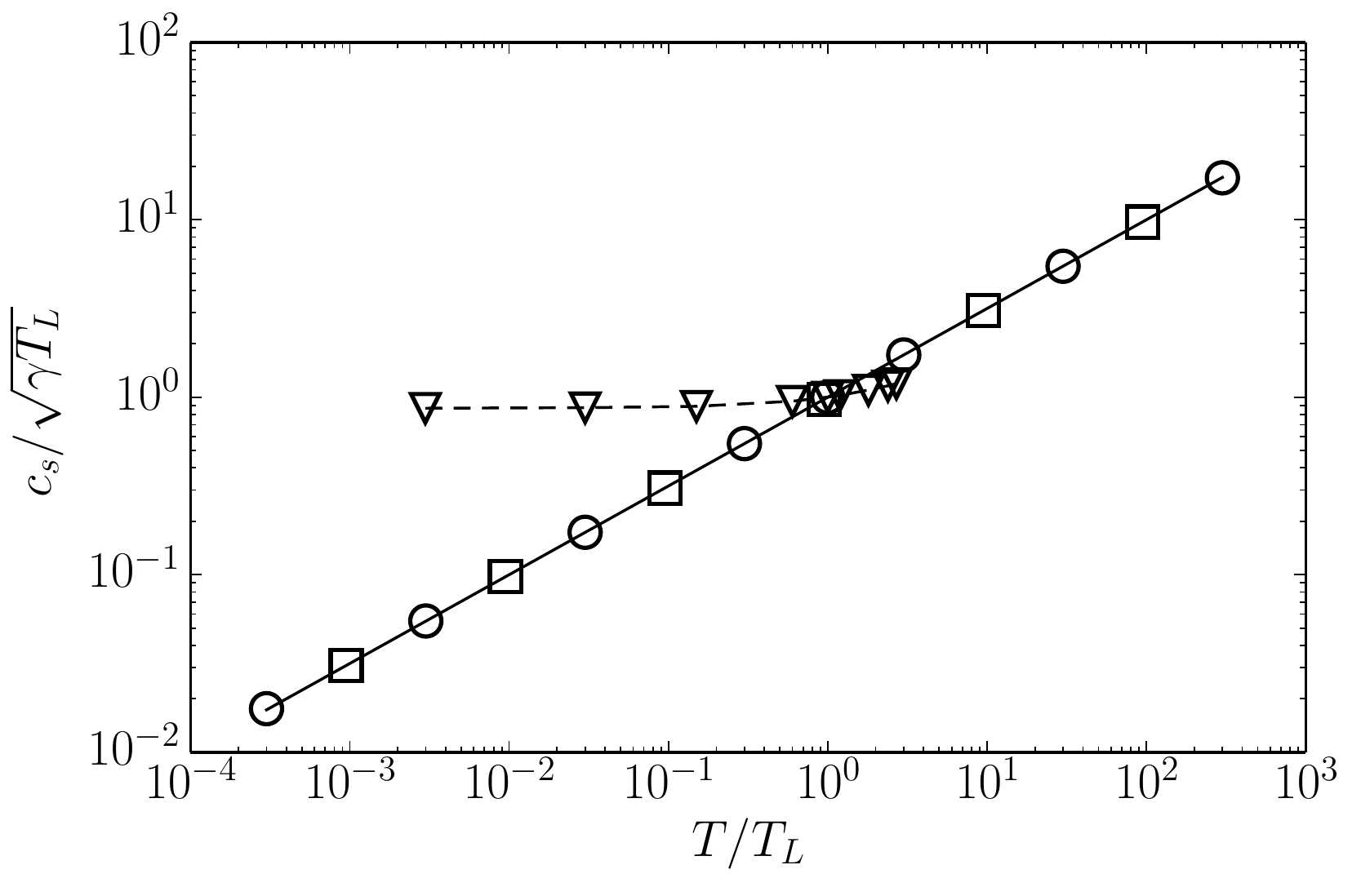}
	\caption{{Speed of sound with $D2Q9$. 
		Line: theory, $c_{\rm s}=\sqrt{2T}$; 
        Circle: present method without advection;
        Square: present method with {an advection} Mach number $\rm{Ma}_{\rm a}=10$. 
        Triangle: thermal LBGK \cite{Prasianakis2009}.}}
	\label{fig:cs}
\end{figure}

\begin{figure}[t]
	\includegraphics[width=0.45\textwidth]{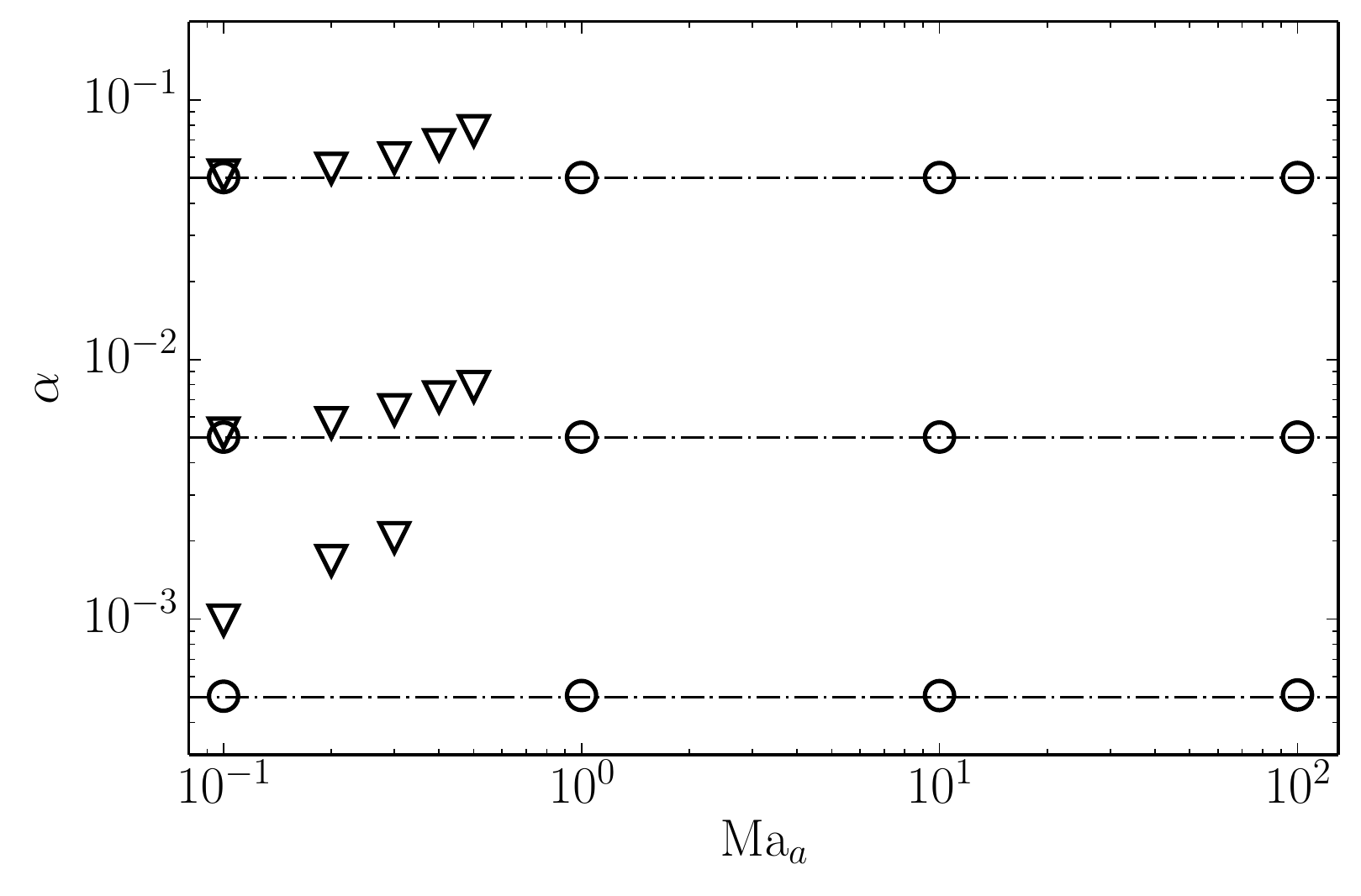}
	\caption{
      Thermal diffusivity with $D2Q9$ lattice at various advection Mach numbers ${\rm Ma}_a = u/\sqrt{2T}$. 
      Line: theory; Circle: present method; Triangle: thermal LBGK \cite{Ansumali2005,Prasianakis2009}.}
	\label{fig:alpha}
\end{figure}

\noindent
We now turn to the compressible flow while still using the nine-speeds $D2Q9$. The difference with the above isothermal model is that now the energy conservation (\ref{eq:conservation_T}) is included in the predictor-corrector propagation step of the algorithm.
The LBM counterpart is the thermal LBGK \cite{Ansumali2005}.
{The} first numerical experiment concerns measuring the speed of sound and comparing it to the theoretical prediction, 
$c_{\rm s}=\sqrt{\gamma T}$, where the adiabatic exponent $\gamma=2$ for two-dimensional ideal gas. 
To that end, speed of sound was measured by introducing a pressure disturbance $ \Delta p =10^{-3}$ and
tracking the resulting shock front. Results for a fluid at rest, and advected with $\rm{Ma}_a=10$ are presented in Fig.\ \ref{fig:cs}.  
It is apparent that the speed of sound measured in the simulation excellently agrees with theory
for all temperatures in the range $T \in [10^{-4}, 10^2]$, 
irrespectively of the advection speed.  
Fig.\ \ref{fig:cs} also 
shows that the thermal LBGK with nine speeds  matches the correct speed of sound only {\it at} the lattice temperature $T=T_{\rm L}$ \cite{Prasianakis2009}.  Thus, the present method extends the physical relevance of thermal LBGK by about six decades in terms of temperature range.
We further probe the conduction of heat by measuring thermal diffusivity from the decay of a sinusoidal temperature profile \cite{Nie2008}.
A periodic 
set-up is chosen with an initial 
density $\rho  = A \sin(2 \pi x/L) + \rho_0$
at
constant pressure $p = \rho_0 T_L$, with amplitude $A=0.001$, $\rho_0=1$ and
longitudinal resolution $L=300$. 
Theoretical prediction of thermal diffusivity for the $D2Q9$ model is $\alpha=(T/4)(1/2\beta - 1/2)\delta t$ \cite{Prasianakis2009}.
Fig.\ \ref{fig:alpha} demonstrates
excellent agreement between theory and numerical results, for a range
of advection speed up to ${\rm Ma}_{\rm a} = 100$, whereas thermal LBGK \cite{Ansumali2005} shows severe deviations.

\begin{figure}[t]
    \includegraphics[width=0.45\textwidth]{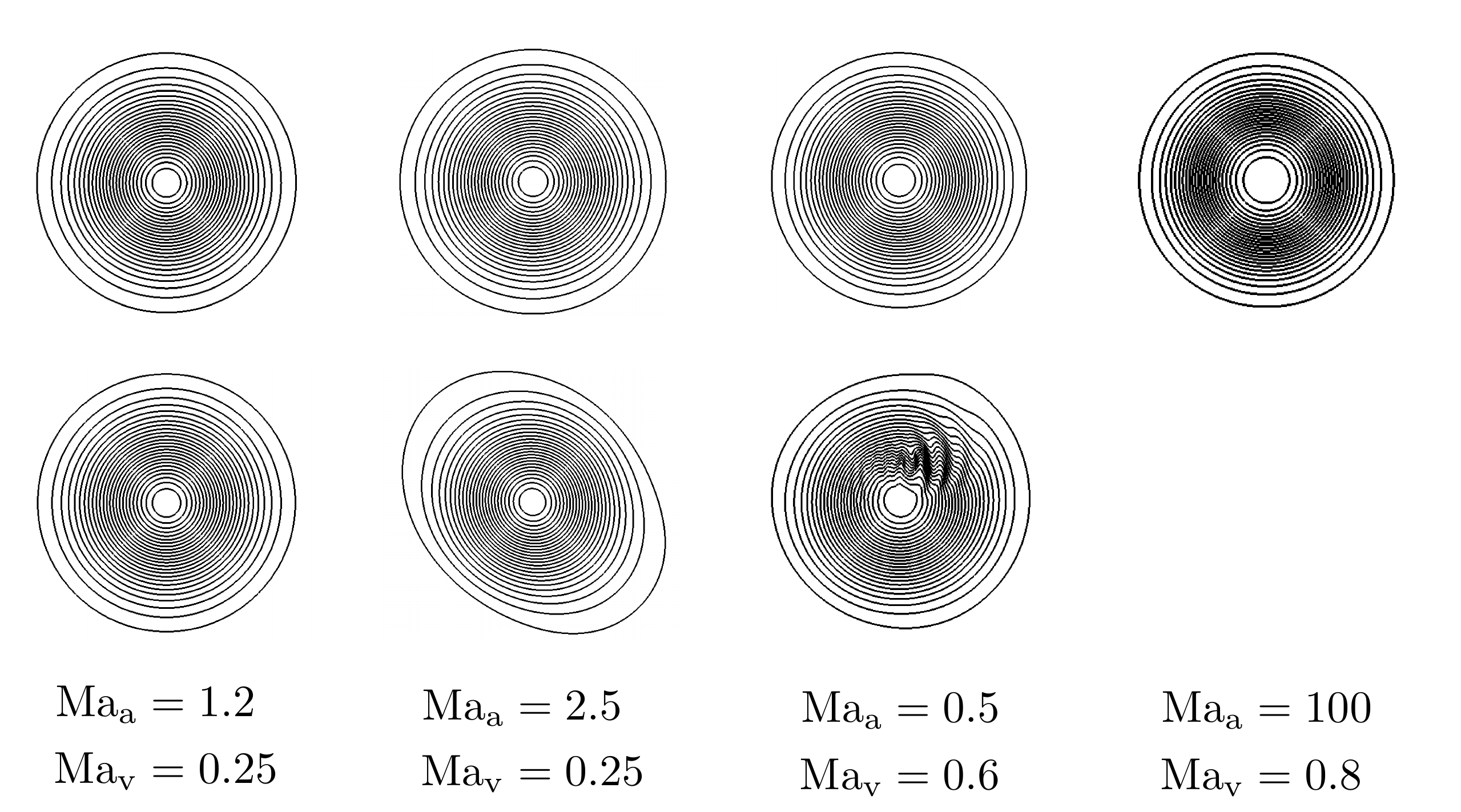}
	\caption{{Pressure contours of the vortex propagation for various advection and vortex Mach numbers. 
	         Top row: present method, $D2Q9$; Bottom row: entropic LBGK, $D2Q49$ \cite{Frapolli2016}.}}
	\label{fig:advection}
\end{figure}

\noindent In general, simulations of compressible flows with LBM require higher-order lattices, with a much larger number of discrete speeds \cite{Frapolli2016,Shan2016,Sagaut2017,Philippi2017}. We conclude this Letter by comparing the above nine-speeds $D2Q9$ model with the entropic LBM on a higher-order lattice with {forty-nine} speeds, $D2Q49$ \cite{Frapolli2016}.
The benchmark consists {of} the advection of a vortex by a uniform flow.
The vortex {with radius $R$} is propagated with advection Mach
number $\rm{Ma}_{\rm a} = U_\infty / \sqrt{2 T_\infty}$ while the
vortex Mach number $\rm{Ma}_{\rm v}$ defines the tangential velocity of the vortex
$u_\varphi (r)= {\rm{Ma}}_{\rm v} r \exp [ (1-r^2)/2 ]$, where $r=r'/R$ is the reduced
radius \cite{Taylor1918,Inoue1999}.
{In Fig.\ \ref{fig:advection}, pressure contours are shown for the
present $D2Q9$ model (top row), together with those computed by the entropic LBM $D2Q49$ \cite{Frapolli2016} (bottom row),
for various combinations of $\rm{Ma}_{\rm a}$ and $\rm{Ma}_{\rm v}$.
Note that LBM \cite{Frapolli2016} is in a global gauge $\lambda = \{ { \bm U }, T_{\rm
L}\}$, ${\bm U}=(1,0)$; this minimizes errors whenever ${u}_x\sim 1$.
Clearly, with a global gauge conveniently chosen,
unidirectional advection at small vortex Mach numbers can be  accomplished
with LBM (Fig.\ \ref{fig:advection}, first column).  
However, deviations of the local velocity and
temperature away from the global gauge eventually lead to spurious deformation
of the vortex (Fig.\ \ref{fig:advection}, second and third column).  
In contrast, present method shows no  deformation of the propagating vortex, even for large Mach numbers (Fig.\ \ref{fig:advection}, last column).
This shows superiority of the present method over the higher-order LBM.}\\
Other pertinent aspects were studied using this benchmark. We observed that the predictor-corrector
	tailoring of the particles 
	required about two to three iterations to convergence, with maximum of five at a fraction of grid points, when the gauge was initialized  as in  (\ref{eq:seed_gauge},\ref{eq:seed_gauge_T}); see Appendix~\ref{app:predictorcorrector}.
	Independence of the limit 
	from the seed gauge was probed by choosing different values of $\bm{u}_0$ and $T_0$; for example $\bm{u}_0=\bm{0}$, $T_0=T_{\rm L}$, or even \textquotedblleft unnatural" $\bm{u}_0=-\bm{u}(\bm{x},t-\delta t)$.
%(the flow direction is suddenly reverted).
We found that converged values are independent of the initialization which reveals that flow density, velocity and temperature are indeed defined correctly by 
%Eqs.\ (\ref{eq:density}), (\ref{eq:flow}) and (\ref{eq:T}).
the limits\ (\ref{eq:density}), (\ref{eq:flow}) and (\ref{eq:T}).
% (that is, they are independent of a particular sequence of iterations leading to the limit).  	

\noindent 
Thus, we can view the particles as an attractor of the predictor-corrector process. 
Basin of attractor depends on the  Mach number and narrows down at larger values;
however, the seed gauge (\ref{eq:seed_gauge},\ref{eq:seed_gauge_T}) was always included in the basin. 
This shows robustness of emerging kinetic picture.

\newpage

\noindent Summarizing, the LBM is rigorously valid in the limit of
vanishing flow velocity and at fixed lattice temperature. Practitioners of LBM circumvent these limitations
by setting empirical bounds on the allowed variation of velocity and
temperature (e.\ g., the flow velocity to stay below ten percent of
the lattice speed of sound, $\|\bm{u}\|\lesssim 0.1 \sqrt{T_{\rm L}}$, 
a common recommendation for incompressible flow simulations;
see Fig.\ \ref{fig:visc}). However, such heuristic constraints cannot be
universally maintained and quickly become meaningless especially for compressible flows.  \\
In this Letter, we proposed a major revision of the kinetic theory for fluid dynamics by constructing \textquotedblleft particles-on-demand" instead of  a priori fixed.
Its realization demonstrates that the range of accessible flow velocities and temperatures
becomes eventually unlimited.  
Same as in the LBM, the collision step retains locality and makes 
application of advanced collision models, already elaborated in LBM, straightforward in the present context, e.\ g.\ for varying Prandtl number and adiabatic exponent.
The new discrete kinetic theory necessarily abandons the LBM lattice propagation since tailoring particles' velocities does not match to the links of a lattice. While the propagation step becomes computationally more intensive than in LBM, the algorithm is still fully explicit, and, as our simulations show, the net demand is lower than that of the higher-order LBM while the operation domain is incomparably larger. Finally, error-free collision results in outstanding numerical stability even with the simplest BGK model. 
This all, as we believe, opens up an entirely new perspective on complex flow simulations.

\noindent
This work was supported by the SNF grants P2EZP2\_178436 (B.D.) and 200021-172640 (F.B.), and the ETH research grant ETH-13 17-1.
Computational resources at the Swiss National Super Computing Center (CSCS)
were provided under the grant s800.

\newpage

%\bibliography{apssamp}
\providecommand{\noopsort}[1]{}\providecommand{\singleletter}[1]{#1}%

\appendix

\section{Equilibrium}
\label{app:equilibrium}

We consider the standard nine-velocity model, the $D2Q9$ lattice. The discrete speeds are constructed as a tensor product of two one-dimensional peculiar speeds, 
$c_{i}=i$, 
where $i=0,\pm 1$. Discrete speeds in two-dimensions are
\begin{equation}
{\bm c}_{(i,j)} = (c_i, c_j)^\dagger,
\end{equation}
where we have introduced two-dimensional indices in order to reflect the Cartesian frame instead of a more common single subscript.
Thus, the discrete velocities are defined as
\begin{equation} \label{eq:vel}
{\bm v}_{(i,j)} = \sqrt{\theta} 
\begin{pmatrix}
c_i \\ 
c_j
\end{pmatrix} +
\begin{pmatrix}
u_x \\ 
u_y
\end{pmatrix},
\end{equation}
with reduced temperature $\theta = T/T_L$ and lattice temperature $T_L = 1/3$.
Populations are labelled as well with two indices, $f_{(i,j)}$,  
corresponding to their respective velocities~\eqref{eq:vel}.
The local equilibrium populations are now conveniently expressed as the product of one-dimensional weights
\begin{equation}
f^{\rm eq}_{(i,j)} = \rho W_{(i,j)} = \rho W_i W_j,
\end{equation}
where
\begin{equation}
W_i = \left \lbrace \begin{array}{ll}
        2/3, & \text{for } i=0, \\
        1/6,      & \text{otherwise}.
\end{array}\right.
\end{equation}
While the equilibrium populations are constant up to the proportionality to density, their moments 
\begin{equation}
M^{{\rm eq}}_{mn} = \rho  \sum_{(i,j)} W_i W_j (\sqrt{\theta}{c}_{i}+{u}_x)^m(\sqrt{\theta}{c}_{j}+{u}_y)^n,
\end{equation}
recover the pertinent Maxwell-Boltzmann moments up to the fourth order, $m+n = 4$, without error for any
temperature and velocity.

\section{Transfer Matrix}
\label{app:transfer}

Populations $f_{(i,j)}^\lambda$ measured in the gauge $\lambda$, can be represented as 
linear combinations of $9$ linearly independent moments,
\begin{equation}
M^\lambda = 
%\begin{split}
(M^{\lambda}_{00}, M^{\lambda}_{10}, M^{\lambda}_{01}, M^{\lambda}_{11}, M^{\lambda}_{20}, M^{\lambda}_{02}, 
M^{\lambda}_{21}, M^{\lambda}_{12}, M^{\lambda}_{22})^\dagger,
%\end{split}
\end{equation}
see also Eq.~\eqref{eq:moments} in the main text,
and $\mathcal{M}_{\lambda}$ is the $Q\times Q$ matrix 
of the linear map between populations and moments,
\begin{equation}
\mathcal{M}_{\lambda}f^\lambda = M^\lambda.
\end{equation}
Moments are invariant with respect to the gauge,
\begin{equation}\label{eq:invariance}
%M^{\lambda}_{mn}=M^{\lambda'}_{mn},
\mathcal{M}_{\lambda'} f^{\lambda'} = \mathcal{M}_{\lambda}f^{\lambda},
\end{equation}
and the transfer from gauge $\lambda$ to $\lambda'$ 
can be written in the following explicit form,
\begin{equation}\label{eq:transform2}
%f'_{(\alpha,\beta)} = \frac{\omega(\alpha)\omega(\beta)}{\left(3T'\right)^2} \sum_{\gamma,\delta} g_x(\gamma,\alpha) g_y(\delta,\beta) 
f_{(k,l)}^{\lambda'} = \omega(k)\omega(l) \sum_{i,j} g_x(i,k) g_y(j,l) f_{(i,j)}^\lambda,
\end{equation}
where
\begin{align}
g_{\xi}(i, j) &= A_{\xi}^2(i) - B_{\xi}(i,j), \\
A_{\xi}(i) &= \left(u_{\xi}' - u_{\xi}\right)/\sqrt{3} - i \sqrt{T}, \\
B_{\xi}(i,j) &= 
\left \lbrace \begin{array}{rl}
        T,                     & \text{for } j=0, \\
        j \sqrt{T} A_{\xi}(i), & \text{otherwise}, 
\end{array}\right. \\
\omega(i) &= 
\left \lbrace \begin{array}{rl}
        1/T', & \text{for } i=0, \\
        \quad -1/2 T', & \text{otherwise}.
\end{array}\right.
\end{align}
Formula~\eqref{eq:transform2} only involves evaluation of a dot-product as opposed to numerically solving the linear system~\eqref{eq:invariance}.

\begin{figure}[t]
\includegraphics[width=0.5\textwidth]{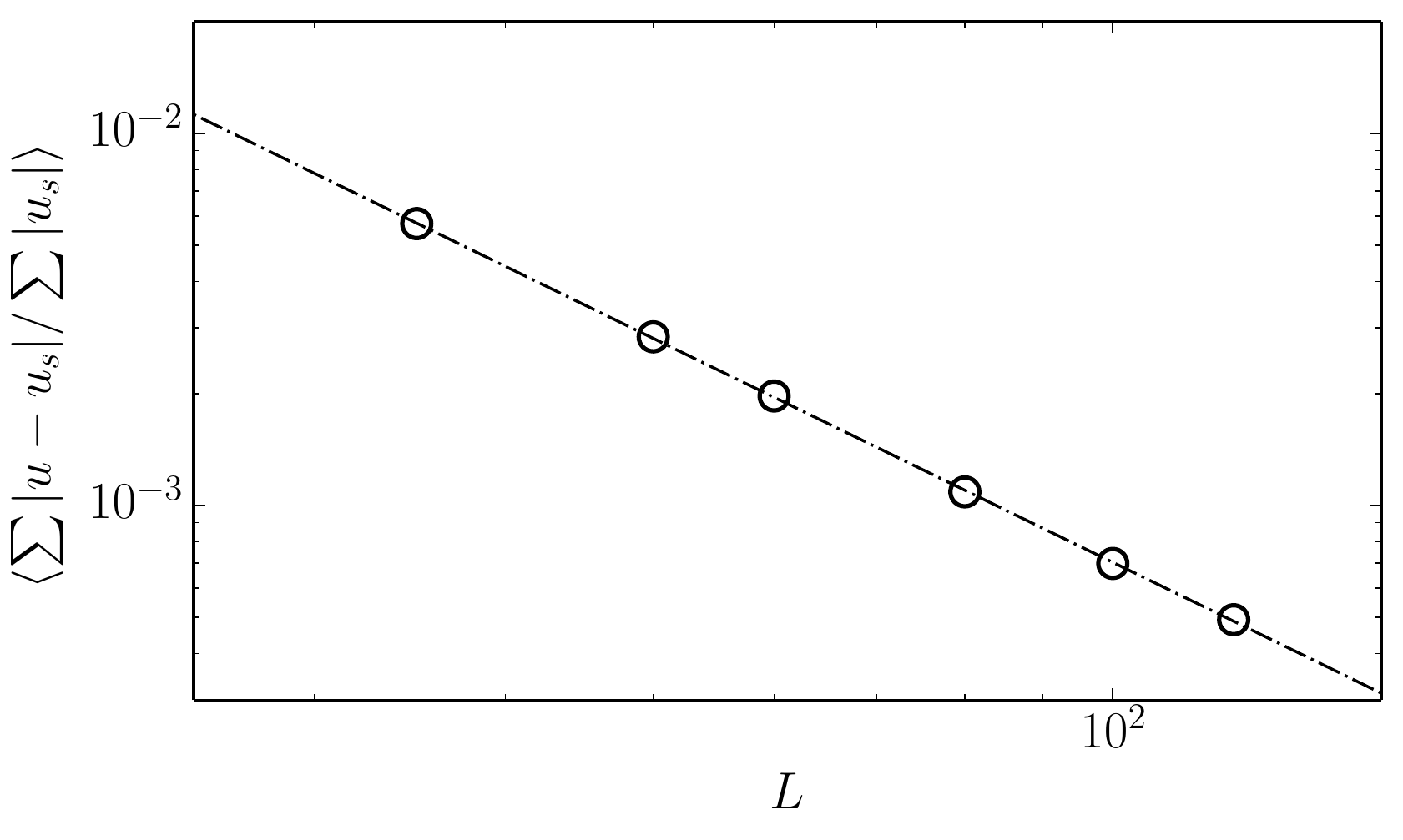}
\caption{Convergence rate of Green-Taylor vortex flow for grid resolutions $L \in [35,120]$ and
Reynolds number ${\rm Re} = u_0 L/\nu = 50$. Symbols: relative error with respect to solution ${\bm u}_s$; Line: second order convergence. }
\label{fig:convergence}
\end{figure}

\section{Reconstruction}
\label{app:reconstruction}

An equidistant rectilinear mesh with $\Delta x = 1$ is used for all simulations. 
Populations at off-grid locations
are reconstructed using $3^{\rm rd}$-order polynomial interpolation,
\begin{equation}
%\sum_{m = 0}^3 \sum_{n = 0}^3 \ell_m \ell_n f_{(i,j)}((x_0 + n, y_0 + m), t)
%\tilde{f}^{\lambda'}_{(i,j)}(\bm{x},t) = \sum_{\substack{0 \leq m \leq 3 \\ 0 \leq n \leq 3}} \ell_{mn}({\bm x}) 
\tilde{f}^{\lambda}_{(i,j)}(\bm{x},t) = \sum_{\substack{0 \leq m \leq 3 \\ 0 \leq n \leq 3}} a_{mn}({\bm x})
f^{\lambda}_{(i,j)}\left((x_0 + n, y_0 + m), t\right),
\end{equation}
where the populations at integer collocation points 
$(x_0 + n, y_0 + m)$ are transformed to gauge $\lambda$ using eq.~\eqref{eq:transform}
and $a_{mn}$ are standard Lagrange polynomials,
\begin{equation}
%\ell_{mn}({\bm x}) = 
a_{mn}({\bm x}) = 
\prod_{\substack{0\leq k \leq 3\\k \neq n}} \frac{ (x - x_0) - k}{n-k} 
\prod_{\substack{0\leq l \leq 3\\l \neq m}} \frac{ (y - y_0) - l}{m-l},
\end{equation}
with respect to reference coordinate,
\begin{equation}
{\bm x}_{0} = (\lfloor x \rfloor - 1, \lfloor y \rfloor - 1),
\end{equation}
where the operation $\lfloor \varphi \rfloor$ rounds down to the largest integer value not greater than $\varphi$.

\begin{figure}[t]
\includegraphics[width=0.44\textwidth]{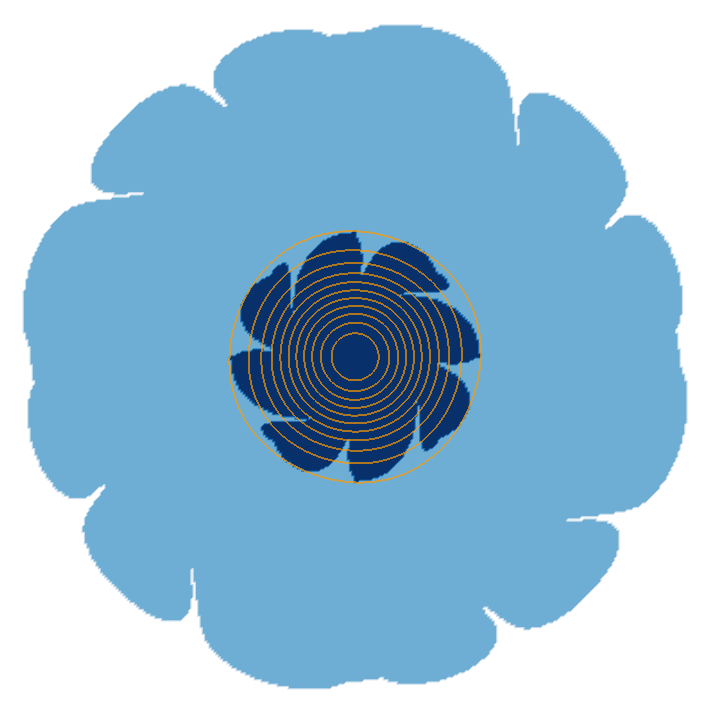}
\caption{Instantaneous recording of a standing vortex with rotation Mach number ${\rm Ma}_v = 0.8$. Colors indicate number of iterations
required for convergence of predictor-corrector scheme (white: 1, light blue: 2, dark blue: 3). Lines: density contours.}
\label{fig:iterations}
\end{figure}

\section{Convergence Order}
\label{app:convergence}

Convergence with respect to grid resolution of the present method was tested 
using the well known periodic Green-Taylor vortex flow.Analytical solution of the flow field is given by
\begin{align}
u_x({\bm x},t) &= -(u_0/\sqrt{2}) \cos(k x) \sin(k y) \exp(-2 \nu k^2 t), \\
u_y({\bm x},t) &=  (u_0/\sqrt{2}) \sin(k x) \cos(k y) \exp(-2 \nu k^2 t),
\end{align}
with wave number $k = 2 \pi/L$ and domain size $L$.
In order to maintain
incompressibility,
a small characteristic Mach number
${\rm Ma} = u_0/\sqrt{T}  = 0.001 $ was chosen and the simulation was run at
isothermal conditions $T = 3 T_L = 1$. Thus, the speed of sound is $\sqrt{3}$ times 
larger than in a standard LBM simulation with the same lattice. Initial density was set
to unity, $\rho_0 = 1$, and simulated flow field is compared to with respect to theoretical prediction. 
Fig.\ \ref{fig:convergence} shows the rate of convergence of the
relative error averaged over a time period $[0.9 t_h, 1.1 t_h]$, where $t_h$ is the half-decay time.
The present scheme recovers second order of accuracy, which coincides with standard LBM and its semi-Lagrangian variant.

\section{Predictor-Corrector Scheme}
\label{app:predictorcorrector}

The number of predictor-corrector iterations depends on the flow and initial seed gauges, however,
unique solution is found independent of the initial guess values.
On average three
iterations lead to convergence, which is defined for iteration $n+1$ of field $\phi$ by
\begin{equation}
\label{eq:convergence}
\left \lvert \phi_{n+1} - \phi_{n} \right \rvert < \epsilon_{\rm abs} + \epsilon_{\rm rel} \phi_{n+1},
\end{equation}
where
absolute tolerance $\epsilon_{\rm abs} = 10^{-12}$ and relative tolerance $\epsilon_{\rm rel} = 10^{-10}$ are used in
the simulations. Convergence criterion \eqref{eq:convergence} must be separately fulfilled for $\phi = \lbrace u_x, u_y. \sqrt{\theta}\rbrace$.
Fig.\ \ref{fig:iterations} shows the number of iterations at a particular instant in time for the 
standing vortex with vortex Mach number ${\rm Ma}_v = 0.8$ (see main text for definition of the flow).
Superimposed density contours indicate the center of the vortex. It is apparent that generally more iterations
are needed in regions where the flow changes rapidly, and thus, initial seed values based on the previous time step 
(Eqs.\ \eqref{eq:seed_gauge} and~\eqref{eq:seed_gauge_T} in the main text) are farther from the converged result. A maximum number of 5 iterations was recorded
for high advection Mach numbers ${\rm Ma}_a \approx 100$.

\end{document}